# Die Zukunft sehen: Die Chancen und Herausforderungen der Erweiterten und Virtuellen Realität für industrielle Anwendungen

Prof. Dr. Jens Grubert
Hochschule Coburg
jg@jensgrubert.de

Die Digitalisierung, d.h. die fortschreitende Umstellung bestehender und Entwicklung neuer (Geschäfts)prozesse durch digitale Technologien bietet für den Wirtschaftsstandort Deutschland neue Chancen, stellt etablierte Unternehmen jedoch auch vor Herausforderungen. Eine Konkretisierung der Digitalisierung manifestiert sich sich im Verbund von Cyber-physischen-Systemen, dem Internet der Dinge und Cloud Computing in der Industrie 4.0. Die Verzahnung der industriellen Produktion mit moderner Informations- und Kommunikationstechnik zielt darauf ab einen (weitgehenden) selbstorganisierten und selbstoptimierten Produktlebenszyklus zu realisieren (von der Idee über die Fertigung bis zur Wartung und Recycling). Neben Aspekten der Betriebswirtschaft (Geschäftsprozesse), der Sicherheit (IT-Security), der Automatisierung (Big Data und künstliche Intelligenz) spielt aber weiterhin die direkte Interaktion zwischen Mensch (Werker oder Verbraucher) und physischen Produkten (Betriebsmittel oder Endprodukt) eine wesentliche Rolle für einen gelingenden digitalen Wandel. Denn letztendlich produzieren immer noch physische Maschinen. Kunden werden auch weiterhin nicht ausschließlich digitale Konsumgüter erwerben sondern weiterhin Produkte zum anfassen und erleben nutzen.

Menschen benötigen geeignete Benutzungsschnittstellen, d.h. Brücken um mit Produkten in Kontakt zu treten, um diese teils digitalen, teils physischen Prozesse und Produkte gestalten, überprüfen, steuern und erleben zu können. Etablierte Benutzungsschnittstellen werden den Anforderungen an Produkte, Prozesse und Dienstleistungen, die einen variablen Anteil an digitalen und physischen Aspekten haben, aber immer seltener gerecht. Die klassische Desktop-PC Benutzungsschnittstelle (Windows, Icons, Menus, Pointer) [MYE98], welche die Interaktion mit Computern seit den 1980er Jahren prägte, richtete sich primär an Wissensarbeiter in Büroumgebungen. Es wurden Interaktionsmetaphern (wie der Schreibtisch oder der Papierkorb) verwendet um mit rein digitalen Information auf artverwandte Weise arbeiten zu können wie mit physischen Akten und Ordnern. Die seit Ende der 2000er Jahre etablierten Benutzungsschnittstellen auf Smartphones oder Tablets ermöglichen die Interaktion mit Mikroaufgaben [OTR05] in mobilen Kontexten. Aber auch hier ist der Fokus auf rein digitalen Informationen (z.B. der Status von Bekannten auf Facebook, Aktienkurse oder Wetterinformation). Aber, ob in Zukunft z.B. alle Informationen über vernetzte Autos ausschließlich über das Smartphone von Nutzern abgerufen werden sollten, oder ob komplexe Produktionsmaschinen rein durch ein Tablet, ohne jeglichen physischen Eingriff, gewartet werden können, darf bezweifelt werden.

Stattdessen sollten neuartige Benutzungsschnittstellen auf den spezifischen Charakter von physischen und digitalen Anteilen an einem Produkt oder Produktionsmittel Rücksicht nehmen. Das Gebiet der vermischten Realität (engl. Mixed Reality) mit seinen Ausprägungen der erweiterten Realität (engl. Augmented Reality, AR) und der virtuellen Realität (Virtual Reality, VR) zielt genau darauf ab. Die Erweiterte Realität ermöglicht es digitale Informationen in räumlichen Zusammenhängen in physischen Umgebungen erlebbar zu machen, siehe Abbildung 1. Die Virtuelle Realität hingegen zielt darauf ab reale (oder erdachte) Umgebungen durch 3D Computergrafiken bestmöglich zu simulieren und Benutzer in diese Welten immersiv eintauchen zu lassen, siehe Abbildung 2.

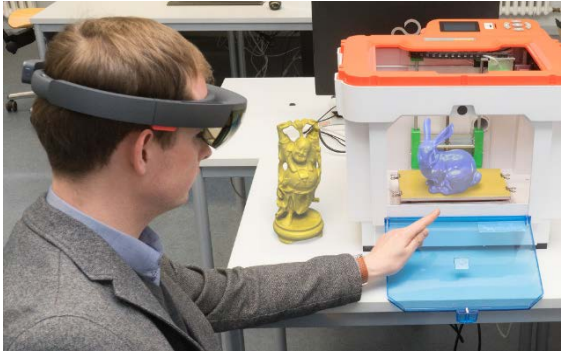

Abbildung 1: Eine Datenbrille zeigt mittels Erweiterter Realität virtuelle Objekte (Hase, Budda-Statue an) bevor diese physisch 3D-gedruckt werden.

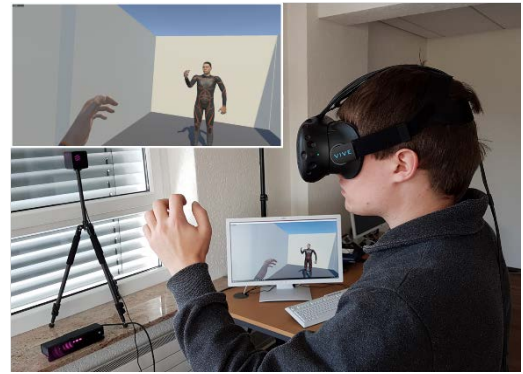

Abbildung 2: In der virtuellen Realität befinden sich Nutzer in simulierten und immersiven 3D Welten, hier eine Person vor einem Spiegel in einer medizinischen VR Anwendung.

Beide befinden sich auf verschiedenen Enden eines Kontinuums, welches als Mixed Reality bekannt ist und physische mit digitalen Inhalten zu verschiedenen Anteilen vermischt [MK94]. Visionen, reale und virtuelle Welten zu vermischen, bestehen schon lange. Die Virtuelle Realität hat ihre geistigen Vorläufer schon in Plato's Höhlengleichnis oder Descartes' Genius malignus [STE16]. Die Erweiterte Realität wurde in Frank L. Baums 1901 erschienenen Novelle "The Master Key" schon früh beschrieben. Hier wird ein "Character Marker" verwendet, eine Brille, welche durch "elektrische Vibrationen" die Persönlichkeit einer gegenüberstehenden Person an deren Kopf durch das Einblenden eines Buchstaben sichtbar macht. Seit den 1960er Jahren haben Forscher und Entwickler daran gearbeitet diese Visionen technisch zu ermöglichen. So schuf bereits im Jahr 1968 Ivan Sutherland eine erste Datenbrille, welche im Raum verortete dreidimensionale Daten anzeigen konnte und es gleichzeitig erlaubte die physische Umgebung wahrzunehmen [SUT68]. Zur gleichen Zeit setzten unter Führung von Tom Furness jahrzehntelange Entwicklungen bei der U.S. Airforce ein, ein "Super-Cockpit" zu erschaffen, welches durch die Einblendung virtueller Informationen Piloten bei Ihren Aufgaben besser unterstützen sollte. Eines der ersten industriellen Einsatzszenarien der Erweiterten Realität wurde in den frühen 1990er Jahren bei Boeing realisiert [CM92].

Hier sollten Werker mittels Datenbrillen bei der Verlegung von Kabelbäumen in Flugzeugen unterstützt werden. Das System setzte sich aber unter anderem aufgrund ergonomischer Probleme (u.a. zu schwere Datenbrille) und des komplizierten Erstellens der digitalen Vorlagen nicht durch [WON16]. Auch die virtuelle Realität wurde in den 1990er Jahren erstmals in industriellen Kontexten genutzt. So wurden bei British Airways Flugsimulatoren eingesetzt oder bei Daimler-Chrysler Ergonomieuntersuchungen in virtuellen Fahrkabinen durchgeführt [BRO99]. Dennoch konnten sich diese Technologien damals nicht sehr weit verbreiten. Neben technischen Einschränkungen (z.B. Verzögerung, geringe Anzeigeauflösung, niedrige Verortungsqualität, nur stationärer Betrieb mit sehr hohem Platzaufwand) waren auch die Kosten solcher Systeme für viele Unternehmen zu hoch. So kostete damals ein British Airways Flugsimulator ca. 13 Millionen USD. Selbst einfach desktop-basierte Systeme lagen in der Größenordnung von 150.000 USD. Auch die Einführung verbraucherorientierter Produkte Mitte der 1990er Jahre (wie Virtual Reality Spielhallen oder der Nintendo Virtual Boy) setzte sich nicht durch.

Mehr als 20 Jahre später lohnt sich jedoch ein neuer Blick auf diese Technologien, da sich die Rahmenbedingungen für deren Einsatz stark gewandelt haben. Unternehmen wie Microsoft, Facebook oder HTC investieren seit einigen Jahren massiv in die virtuelle und erweiterte Realität. So übernahm Facebook im Jahr 2014 für ca. 2 Milliarden Euro den damals kleinen Hersteller von VR-Datenbrillen Oculus. Mittlerweile kann für wenige Euro jedes Smartphone in eine Datenbrille für die Virtuelle Realität umfunktioniert werden. Die technisch anspruchsvollen Datenbrillen für die Erweiterte Realität befinden sich auch kurz vor dem Eintritt in den verbraucherorientierten Markt. Kostete eine einfache Datenbrille vor 10 Jahren noch ab 10.000 Euro aufwärts (ohne Recheneinheiten oder Verortungssysteme) so sind heute bereits in sich geschlossene kopfgetragene stereoskopische Datenbrillen mit integriertem PC und überzeugender Verortungstechnologie zwischen 800 und 3.000 Euro zu erwerben[1]. Weiterhin unterstützt der immer weiter voranschreitende Ausbau der Telekommunikationsinfrastruktur (Einführung von LTE in Deutschland seit 2010, voranstehende Einführung von 5G ab 2018) die schnelle Übertragung von digitalen Informationen, so dass auch große Datenmengen, die teilweise für Mixed Reality Anwendungen benötigt werden, sehr schnell übertragen werden können. Weiterhin wird die Erstellung von virtuellen Inhalten z.B. durch den Gebrauch von leistbaren 360° Kameras vereinfacht. Durch diese Entwicklungen lassen sich nun auch für kleinere und mittelständige Unternehmen innovative Lösungen mittels Mixed Reality umsetzen.

In der Konzeptionierung und Gestaltung von Produkten werden vor allem Lösungen im Bereich der VR immer beliebter, da der Einsatz dieser Technologie zu kürzeren Iterationszyklen im Design und zu schnelleren Gestaltungsentscheidungen führen kann [CJN15]. Dies reicht von einzelnen Werkstücken bis zur Planung von ganzen Fabriken [MYW12]. Darüber hinaus kann die Erweiterte Realität auch in bestehenden Fabrikhallen eingesetzt werden um so z.B. die Einrichtung von Fertigungsbereichen vor Ort zu planen [DSA03] oder Werker bei Kommissionierungsaufgaben zu utnerstützen [GHM10]. Auch die virtuelle Inbetriebnahme von komplexen Produkten hält zunehmend Einzug in Unternehmen und ist mittlerweile in einer VDI/VDE Richtlinie gemündet [VDI16]. Hier können z.B. reale Anlagensteuerungsbefehle mit virtuellen Maschinen verknüpft werden um so frühzeitig den Aufbau von Fertigungsstraßen zu optimieren.

Die Unterstützung der Kommunikation zwischen örtlich entfernten Gesprächspartnern ist ein weiteres vielversprechendes Anwendungsgebiet. So zielt der Bereich der Telepräsenz darauf ab den Eindruck zu erwecken, als ob entfernte Gesprächspartner so an einem Gespräch teilnehmen als wären sie vor Ort. In Filmen wie Starwars oder Kingsman werden solche Treffen schon lange gezeigt. Mit vorhandenen Video-Chat Lösungen wie Skype, FaceTime oder Hangouts sind solche Treffen bisher jedoch nur bedingt möglich, insbesondere in Situationen mit bis dahin unbekannten Teilnehmern (z.B. im Bereich der Kundenakquise) oder bei der Lösung komplexer Aufgaben [DMJ12]. Technische Entwicklungen ermöglichen es aber mittlerweile Stück für Stück leistbare immersive Telepräsenzsysteme zu erstellen, bei denen eine entfernte Person mit ihrem gesamten Körper an einer Besprechung teilnehmen kann [PKB16, ORF16]. Mittels der VR lassen sich Treffen auch komplett in immersiven Räumen abhalten und so z.B. direkt komplexe räumliche Sachverhalte wie die 3D Konstruktion eines Autos diskutieren. Herausforderungen bestehen hier aber zur Zeit noch in der zufriedenstellenden Darstellung von virtuellen Avataren, da das Gesicht des Nutzers meist von einer VR-Datenbrille verdeckt wird. Erste Ansätze zielen auf eine Rekonstruktion von verdeckten Gesichtsteilen

---

[1] Beispiele Stand April 2017: Epson Moverio BT-300 für 849 Euro, Microsoft Hololens: 3.300 Euro.

[TZS16] oder auf abstrahierte virtuelle Avatare[2] ab.

Weiterhin erlaubt die Kombination von Datenbrillen und schnellen Telekommunikationsnetzen die Unterstützung von Servicetechnikern vor Ort durch Tele-Experten. So können komplexe Wartungs- oder Reparaturvorgänge mittels Fernwartung durchgeführt werden, ohne dass die Experten anreisen müssen. Dadurch lassen sich Stillstandzeiten von Produktionsanlagen stark reduzieren und so Kosten sparen. Auch ohne direkten Kontakt zu einem menschlichen Experten lassen sich Mehrwerte schaffen. So ist es möglich interaktive Schritt-für-Schritt Anleitungen für Wartungen oder Trainingszwecke zu erstellen und somit Kosten in der Ausbildung zu senken oder zeitkritische Aufgaben schneller zu erledigen. Ein Beispiel dafür ist die Rescue Assist Applikation von Daimler für Smartphones und Tablets, mit welcher Ersthelfer bei einem Unfall wichtige im Autoinneren verborgene Fahrzeugteile für alle Mercedes-Benz Modelle seit 1990 mittels AR und VR anzeigen können[3].

Im Bereich des Marketings und Vertrieb etablieren sich mittlerweile verschiedene Mixed Reality Ansätze. Unternehmen wie Ikea oder Lego setzen darauf ihre Produkte nicht nur in Bildern auf gedrucktem Papier oder auf Webseiten zu präsentieren sondern Nutzern interaktive 3D Modelle auf ihren Smartphones zu präsentieren[4]. Zu dem werden Smartphone Applikationen immer beliebter, welche die Produktkonfiguration und -platzierung vor Ort ermöglichen. Beispiele dafür sind Möbel[5] (Platzierung von Schränken, Couches) oder Elektronik (z.B. Fernseher[6]). Lösungen im Bereich VR umfassen u.a. das Produkterlebnis und die Produktkonfiguration von Autos in virtuellen Showrooms[7]. Eine weitere Anwendung besteht in der Überprüfung von Dokumenten mittels AR [HGS13].

Wie oben bereits erwähnt, sinken die Kosten für VR und AR-Hardware und ermöglichen somit leistbare Systeme für kleine und mittelständische Systeme. Steht damit dem Siegeszug dieser Technologien nichts mehr im Wege? Bei genauerer Betrachtung bestehen auch weiterhin Hürden, die eine enge Kooperation zwischen Unternehmen und anwendungsorientierten Forschungseinrichtungen erfordern, um eine nahtlose Integration von Mixed Reality Technologien in bestehende oder zur Schaffung neuer Wirtschaftsprozesse zu gewährleisten.

Während die Qualität von Mixed Reality Hardware steigt und die Kosten sinken (z.B. für Datenbrillen, Verortungssysteme, 360° Kameras), sind die Integration in bestehende Arbeitsprozesse, die Erstellung von sachgerechten 3D Inhalten, die Nutzbarkeit und Konfigurierbarkeit durch Nicht-Programmierer, die Vielfältigkeit einsetzbarer Endgeräte und Plattformen sowie die fehlende Standardisierung von Interaktionsmöglichkeiten Hürden, die überwunden werden müssen.

Im Maschinenbau sind Product-Lifecycle-Management Systeme (wie Siemens NX) unerlässlich um von der digitalen Planung zur physischen Konstruktion zu gelangen. Dazu setzen gängige CAD/CAM Systeme auf proprietäre Dateiformate, die das Visualisieren der 3D Daten in Mixed Reality Anwendungen erschweren. Ein über die Mixed Reality hinausgehender Aspekt sind die Inkompatibilitäten zwischen Systemen für die Produktentwicklung bis Produktionsfreigabe (PLM), Fertigungsplanund und Produktionssimulation (digitale Fabrik),

---

[2] https://www.wired.com/2016/10/oculus-facebook-social-vr/ Letzter Zugriff 13.04.2017.
[3] https://itunes.apple.com/gb/app/rescue-assist/id890940470?mt=8 Letzter Zugriff 13.04.2017.
[4] http://www.augment.com/blog/3-consumer-giants-who-used-augmented-reality-for-retail/ Letzter Zugriff 13.04.2017.
[5] http://www.augmentedfurniture.com/ Letzter Zugriff 13.04.2017.
[6] http://www.samsung.com/global/article/articleDetailView.do?atcl_id=85 Letzter Zugriff 13.04.2017.
[7] https://audi-illustrated.com/en/CES-2016/Audi-VR-experience Letzter Zugriff 13.04.2017.

Manufacturing Execution Systems (MES) und Enterprise Resource Planning (ERP). Diese Inkompatibilitäten erschweren die durch die Digitalisierung angestrebten individualisierte und automatisierte Produktion von Gütern.

Neben der detailgetreuen aber zeitaufwendigen Konstruktion von Bauteilen (durch CAD Systeme) spielen auch Rapid Prototyping Verfahren eine zunehmend größere Rolle für die schnelle und individualisierte Produktion. Insbesondere können durch heute verfügbare Aufnahmesysteme wie in Smartphones integrierte Tiefensensoren (Project Tango, Intel RealSense) oder 3D 360° Kameras einzelne Gegenstände oder ganze Räume zeitnah digitalisiert werden und so tendenziell als Vorlage für individualisierte Produkte verwendet werden. Aber auch hier fehlt es noch an einer durchgehenden Arbeitskette um die digitalisierten Rohdaten (meist als Punktwolke oder grobe Vermaschung) für die Weiterverarbeitung nutzen zu können. Hier könnten Mixed Reality Authoring Werkzeuge helfen, welche eine benutzergeführte Vereinfachung und Anpassung der Daten unterstützen. Bisherige Ansätze Mixed Reality Authoring Systeme für Nicht-Programmierer zugänglich zu machen sind jedoch noch verbesserungsbedürftig. Während sich Visuelle Programmierumgebungen [RMM09] auch im Bereich 3D Spieleprogrammierung durchsetzen[8] und damit auch für die Erstellung von VR-Anwendungen attraktiv werden, gibt es für AR Anwendungen noch keine etablierten Lösungen. Für Anwendungen im Bereich Erweiterte Realität spielt, im Gegensatz zur VR, die Verortung der 3D Informationen in physischen Umgebungen hinzu. Während es erste Ansätze zur Vor-Ort Modellierung und Programmierung (durch Demonstration)[9] gibt, findet man noch kaum Unterstützung bei der Erstellung von AR-Szenen für entfernte Orte.

Nicht zuletzt führen ständig neue Interaktionsgeräte (neben Smartphones auch Smartwatches und Datenbrillen) und Betriebssystemen (Android, iOS, Windows) dazu, dass Nutzer meist kein durchgängiges Benutzungserlebnis erfahren [GKQ15, QG15, GKQ16]. In Zukunft sollten die Vorteile und Nachteile verschiedener Geräteklassen besser verstanden und standardisierte Interaktionsformen entwickelt werden, welche eine kontinuierliche und lückenlose Interaktion mit 3D Informationen in der Mixed Reality über Geräte- (und Betriebssystem)grenzen hinweg ermöglichen. VR-Simulationen [RWB09] von verschiedenen Geräteklassen sowie web-basierte AR- [OGR12] und VR[10]-Plattformen können dabei hilfreich sein.

Zusammengefasst haben in den letzten Jahrzehnten weitreichende Entwicklungen zum Nutzbarmachung von Mixed Reality Technologien stattgefunden. Neben ersten Entwicklungen im Militärbereich wurden schon Anfang der 1990er Jahre die Potentiale dieser Technologie für industrielle Anwendungen erkannt. Durch die Blüte von verbraucherorientierten VR und AR-Produkten seit Mitte der 2010er Jahre ermöglichen sich jetzt neue Chancen diese Technologie als Innovationstreiber auch in kleinen und mittelständischen Betrieben zu etablieren. Damit dies geschehen kann, sollte eine enge Kooperation zwischen anwendungsorientierten Hochschulen und Unternehmen erfolgen um die verbleibenden Herausforderungen erfolgreich bewältigen zu können.

---

[8] z.B. https://docs.unrealengine.com/latest/INT/Engine/Blueprints/ Letzter Zugriff 13.04.2017.

[9] http://www.realityeditor.org/ Letzter Zugriff 13.04.2017.

[10] https://webvr.info/ Letzter Zugriff 13.04.2017.